\documentstyle[epsf,aps]{revtex}
\begin{document}
\title{Report by N. Dinh Dang, V. Kim Au, T. Suzuki, and A. Arima\\ on
the Comment by V. Yu. Ponomarev\thanks{The Comment has been 
later rejected by PRC on January 31, 2002}}
\date{October 24, 2001}
\maketitle
In our recent work~\cite{PDR}, we calculated the 
E1 resonances in neutron-rich oxygen and calcium isotopes within a 
quasiparticle representation of the phonon damping model (PDM) 
in its PDM-1 version~\cite{PDM1a,PDM1b} 
including the superfluid pairing interaction.
It is claimed in the Comment \cite{Pon}
that ``the physical content of the PDM calculations is 
very doubtful'' and the description of the pygmy dipole resonance 
(PDR) is ``not justified'' on ``the quantitative level''. 
However, as seen below, 
the arguments, which the author of the Comment presents to 
prove his case, are either wrong or irrelevant. 

The same Appendix 2D of Ref. [2] in the Comment \cite{Pon} states that 
the assumption of the equal particle-phonon coupling strength 
$F_{ph}^{(q)}=f_{1}$
``may be employed if the width of the strength function is small 
compared to the characteristic energies associated with systematic 
variation in the coupling matrix elements.'' This condition is 
satisfied in the region of E1 resonances considered in \cite{PDR}. 
However, this does not means that the matrix elements of the 
interaction part $V_{q_{1}s_{1}}\equiv\langle 
q_{1}|\sum_{ss'q}F_{ss'}^{(q)}a_{s}^{\dagger}a_{s'}(Q_{q}^{\dagger}
+Q_{q})|s_{1}\rangle$ are the same, as claimed in the Comment \cite{Pon}.

Equation (1) of the Comment \cite{Pon} is not the second
moment for the phonon distribution within the PDM.
The $k$-th moment for the phonon distribution within the PDM is 
calculated (See, e.g.,  Eq. (2.21) of \cite{multi}) as
\begin{equation}
m_{q}^{(k)}=\int_{E_{1}}^{E_{2}}S_{q}(\omega)\omega^{k}d\omega~,
\hspace{2mm} k=1,2,\ldots
\label{mk}
\end{equation}
where $S_{q}(\omega)$ is the PDM strength function
\begin{equation}
S_{q}(\omega)=\frac{\gamma_{q}(\omega)}{(\omega-\bar{\omega})^{2}+
\gamma_{q}^{2}(\omega)}. 
\label{S}
\end{equation}
The energy $\bar{\omega}$ 
of the giant dipole resonance (GDR) is found as the solution of 
Eq. (2.39) in \cite{PDM1b}: 
\begin{equation} 
\bar{\omega}-\omega_{q}-P_{q}(\omega)=0,
\label{pole}
\end{equation}
where ${\omega}_{q}$ is the unperturbed phonon energy (before 
the $ph$-phonon coupling is switched on), and $P_{q}(\omega)$ is the 
polarization operator. The damping $\gamma_{q}(\omega)$ is 
calculated microscopically within PDM as 
the imaginary part of the analytic continuation
of $P_{q}(E)$ into the complex energy plane $E=\omega\pm i\varepsilon$.
Its explicit expression within PDM-1 is given by Eq. (5) of
\cite{PDM1a} (pairing not included) or Eq. (15) of \cite{PDR} (pairing 
included).
There is no way to equalize $m_{q}^{(2)}$ from (\ref{mk}) ($k=$ 2) with
Eq. (1) of the Comment \cite{Pon}. Therefore, all discussions using Eq. 
(1) of the Comment \cite{Pon} with the aim of attaching it to the PDM 
are irrelevant.

The strength function (\ref{S}) is not a Breit-Wigner (BW) 
distribution because the damping $\gamma_{q}(\omega)$ depends on the
energy $\omega$. Such form  
has been derived, for the first time, in \cite{Bogolyubov} using the analytic 
properties of the double-time Green function independently of any
assumption on the coupling matrix elements.
Consequently, the photoabsorption cross section of GDR within the PDM
is not a Lorentzian either. The claim in ~\cite{Pon} 
that a Breit-Wigner 
(Lorentzian) form is assumed or an ad hoc input for the strength function 
(photoabsorption cross section) of GDR within the PDM is simply wrong.

One of the crucial features of the PDM is the use of realistic 
single-particle energies to construct the $ph$ configurations (at zero 
temperature) together with the $pp$ and $hh$ configurations (at 
nonzero temperature) to which 
the GDR is coupled. Therefore a replacement of the realistic 
single-particle spectra with any other ones, such as the random values 
of $E_{s}$ used in the Comment \cite{Pon}, no longer corresponds to the PDM.
So the attempt in \cite{Pon} to imitate the results of the PDM using random
values of $E_{s}$, and an GDR energy $E_{0}$, which is not defined
from (\ref{pole}), is incorrect.

It is by no mean obvious that the coupling constant should increase
as the configuration space gets larger in the same nucleus. 
How the coupling changes is a problem to be 
discussed microscopically. Within the PDM, an increase of the space of $ph$ 
pairs leads to a decrease of the parameter $f_{1}$ to preserve the same
value for the GDR width.

The aim of \cite{PDR} is to use for the calculations of 
E1 resonances in neutron-rich isotopes the same set of two parameters
$(\omega_{q}, f_{1})$, whose values are chosen to reproduce the GDR
in the corresponding double closed-shell nuclei. Therefore, the E1 resonances
in the chains $^{16-24}$O, $^{40-46}$Ca, and $^{48-60}$Ca were 
calculated using the parameters chosen for $^{16}$O, $^{40}$Ca, and 
$^{48}$Ca, respectively. There is no reason why the values of $f_{1}$ 
for $^{40}$Ca and $^{48}$Ca should be the same. The results for GDR 
in $^{16}$O have been obtained already within the enlarged space with
$f_{1}=$ 0.6982 MeV. This value is kept unchanged throughout the chain
of oxygen isotopes as has been mentioned above and in \cite{PDR}. No 
change of the parameter occurs between $^{16}$O and $^{18}$O as 
incorrectly stated in the Comment \cite{Pon}.

The PDM-1 with its two phenomenologically selected parameters allows 
the comparison with the experimental data for only the average 
characteristics of the E1 resonances, such as the overall shape of the cross
section, width, energy, and energy-weighted sum (EWS) of strength. It 
cannot describe such fine structure as the individual low-lying E1 
states measured in \cite{Hartman}. The EWS of E1 strength below 5 MeV
for $^{40}$Ca and $^{48}$Ca are around 0.25 $\%$ and 0.52 $\%$ of the 
Thomas-Reich-Kuhn sum rule (TRK), respectively. They should 
be compared to the experimental 
values of  (0.025$\pm$0.004) $\%$ of TRK 
for $^{40}$Ca, and (0.29$\pm$0.04) $\%$ of TRK for $^{48}$Ca
reported in \cite{Hartman}. The enhancement of strength at low energies
in doubly closed-shell nuclei due to the spreading of GDR has been discussed
in \cite{PDR}. There is no such setting of $B(E1)$ to zero for PDR 
within the PDM as claimed in \cite{Pon}. The problem of 
double counting arises 
only when the structure of phonon is calculated microscopically 
within the random-phase approximation. The use of the 
structureless phonon and the parameters, which are selected
so that the calculated GDR energy reproduces its experimental 
value, excludes any possibility for double counting within the PDM. 

Pairing is not included in the calculations of PDR in Ref. [5] of 
\cite{Pon}.
The aim of Ref. [5] of \cite{Pon} is to see if the PDM is able to
predict the existence of the PDR in neutron-rich nuclei, but not to
reproduce the experimental data. Because of the absence of pairing,
the parameter $f_{1}$ in Ref. [5] of \cite{Pon} 
was increased significantly in the region 
near the Fermi
surface in neutron-rich isotopes. It is natural that
such an increase overestimates the EWS of E1 strength in this region.
This is not the case in the present work \cite{PDR}, where pairing is
included, and the parameters of the model have been chosen to reproduce
the GDR in double closed-shell nuclei.

In conclusion, none of the statements of the Comment \cite{Pon} is relevant. 
The discussions in the Comment \cite{Pon} are fruitless, and its conclusions
are false. The Comment should not be published. 
If, nevertheless, the Editor will accept this 
Comment for publication, 
we would like to submit this report as our Reply to be 
published together with the Comment.
\\~\\
N. Dinh Dang\\
{\it RI-beam factory project office, RIKEN, 
2-1 Hirosawa, Wako, 351-0198 Saitama, Japan}\\
V. Kim Au\\
{\it Cyclotron Institute, Tesax A\&M University, 
College Station, TX 77843-3366, USA}\\
T. Suzuki\\
{\it Department of Physics, Nihon University, Sakurajosui 3-25-40, Setagayaku, Tokyo 156, 
Japan}\\
A. Arima\\
{\it House of Councillors, Nagatacho 2-1-1, Chiyodaku, Tokyo 100-8962, 
Japan}

\end{document}